\documentclass[copyright]{eptcs}
\providecommand{\event}{HSB 2012} 
\usepackage{breakurl}             
\usepackage{graphicx}
\usepackage{subfig}
\usepackage{tikz}

\title{A subsystems approach for parameter estimation of ODE models of hybrid systems}
\author{Anastasis Georgoulas \qquad Allan Clark
\institute{SynthSys--- Synthetic and Systems Biology\\
University of Edinburgh
\\Edinburgh, United Kingdom}
\email{Anastasis.Georgoulas@ed.ac.uk \qquad A.D.Clark@ed.ac.uk}
\and
Andrea Ocone \qquad Stephen Gilmore \qquad Guido Sanguinetti
\institute{School of Informatics\\
University of Edinburgh\\
Edinburgh, United Kingdom}
\email{A.Ocone@ed.ac.uk \qquad Stephen.Gilmore@ed.ac.uk \qquad G.Sanguinetti@ed.ac.uk}
}
\def\titlerunning{A subsystems approach for parameter estimation}
\def\authorrunning{Georgoulas, Clark, Ocone, Gilmore \& Sanguinetti}

\begin{document}
\maketitle

\begin{abstract}
We present a new method for parameter identification of ODE system descriptions based on data measurements. Our method works by splitting the system into a number of subsystems and working on each of them separately, thereby being easily parallelisable, and can also deal with noise in the observations.
\end{abstract}

\section{Introduction}
Kinetic modelling of biochemical systems is a growing area of research in systems biology. The combination of mechanistic insight and predictive power afforded by kinetic models means that these have become a popular tool of investigation in biological modelling. Within the class of kinetic models, ordinary differential equations (ODEs) are by far the dominant modelling paradigm. The importance of nonlinear systems of ODEs stems not only from their value in modelling population data (e.g. microarrays, luciferase assays) but also for their role in describing the average evolution of stochastic discrete and hybrid systems, using tools such as the fluid approximation\cite{6042024} or the linear noise approximation (\cite{1981vanKampen,ssLNA}). The availability of  many analysis tools for ODEs means that qualitative (e.g. the presence of bistability) and quantitative information about the system can readily be obtained from the analysis of the average behaviour of the system.

While the use of non-linear ODEs is an undoubtable success story in systems biology, it is not an unqualified one. The high predictive power of non-linear systems of ODEs often comes at the cost of an explosion in the number of parameters, leading to significant difficulties in calibrating models. Parameter estimation requires a large amount of high quality data even for moderate sized systems. From the computational point of view, this is often exceptionally intensive as it requires solving the system of ODEs many times: typically, the number of solutions scales exponentially with the number of parameters (a particular form of {\it curse of dimensionality}). Therefore, many algorithms for parameter estimation inevitably do not reach convergence, calling into question the validity of model predictions.

Here, we attack the curse of dimensionality of parameter estimation by proposing an approximate solution which effectively reduces a high dimensional problem into several weakly coupled low dimensional problems. Our strategy  builds on the fact that, in many biochemical networks, knowledge of the true trajectory of some species would effectively decouple the parameter estimation problem across different subsystems. Therefore, we propose to use a statistical procedure based on Gaussian Process (GP) regression to infer the full trajectory of each species in the network based on the limited observations. Parameters of each subsystem can then be updated in parallel using a statistically correct Markov Chain Monte Carlo procedure. We show on simulated data sets that this approach is as accurate as the existing state of the art. Furthermore, it is parallelisable, leading to significant computational speed ups, and its statistical nature means that we can accurately quantify the uncertainty in the parameter estimates.

The rest of the paper is organised as follows; we start by introducing some essential statistical concepts which are key ingredients of our approach. We then describe how these concepts can be used in a parameter estimation problem, and present a parallel implementation of the method. We present results on two benchmark data sets, reporting competitive accuracy against state of the art methods. We conclude by discussing the potential and limitations of our approach, as well as indicating avenues for further research. 

\section{Background}

\subsection{Parameter estimation}
There exist a wide variety of parameter estimation methods that have been proposed in the literature and are in use. They all share the notion of exploring the parameter space in order to obtain the optimal values of the parameters, but they can differ significantly in their approach.

The notion of optimality can be expressed through a {\it fitness function}, which expresses how well a certain parameter value can explain the data. Usually, calculating the fitness function involves simulating the behaviour of the system assuming that parameter value is true, and then comparing the results to the observed behaviour. In general, however, the function can reflect one's prior knowledge of the problem or any constraints that are deemed suitable--- for example, by including terms to penalize large values of the parameter. The problem of parameter estimation can then be seen as an optimization problem in terms of the fitness function which we aim to maximize (or, equivalently, an error function which must be minimized).

One point that should be stressed is the importance of the number of parameters. The higher this is, the higher the dimension of the search space and the harder it becomes to efficiently search it. Intuitively, there are more directions in which we can (or must) move in order to explore the search space, therefore the parameter estimation task becomes more complex and time-consuming.

A particularly difficult problem arises when there are multiple parameter sets which can give rise to very similar data. This is further exacerbated by the presence of noise, which means we are often unable to say with certainty what the true values of the data are, making it harder to choose between slightly different results. Recent research~(\cite{C0MB00107D,10.1371/journal.pcbi.0030189}) has shown that wide ranges of parameter sets often produce virtually indistinguishable results, indicating that this is a widespread problem rather than a sporadic one.

\subsection{Markov Chain Monte Carlo}
In this work, we use a Markov Chain Monte Carlo (MCMC) approach, which allows us to selectively and efficiently explore the parameter space (for more details, see, for example, \cite{Gelman95bayesiandata}).

We first assign a {\it prior distribution} $P(p)$ to the parameters, which represents our previous beliefs about their values. For instance, if we have no intrinsic reason to favour one value over another, we can use a uniform distribution as a prior, which would indicate that any parameter value would be equally likely without seeing any data.

However, having some observations introduces additional information which may alter our prior belief. Therefore, the {\it posterior distribution} $P(p\mid D)$ represents the probability of a parameter having the value $p$ after observing the dataset $D$. For example, if we see data which are more likely to have been generated by a specific subset of parameter values, we may start to abandon our uniform prior in favour of a distribution which is biased towards these likelier values. In other words, the posterior distribution represents our belief for the parameter values as determined by the additional knowledge of the observed data.

The MCMC approach can be summarised as follows. Starting with a random set of values $p$ for the parameters, we calculate the posterior probability of the parameters given the data, $P(p\mid D)$. We sample a new set of parameters $p'$ from a Gaussian distribution centred on the current values and repeat the process. The new parameters have probability $\alpha$ of being accepted as a better estimate, where 
$\alpha = \min{(\frac{P(p'\mid D)}{P(p\mid D)},1)}$.
In other words, if the new parameters give rise to a higher likelihood, they are always accepted; otherwise, they can still be accepted with a certain probability. After a number of steps, this procedure converges and we can sample from the posterior distribution of parameters.

This has the advantage of not providing a single-point estimate of the parameters, but rather an entire probability distribution. In practice, we can take a number of samples from the distribution and use them as its representatives. For example, we can perform Kernel Density Estimation\cite{smoothingBowmanAzzalini}, which returns a smoothed histogram approximating the distribution. We can therefore also obtain information about the confidence or uncertainty of the estimated parameter values.

\section{Gaussian Processes regression}
Our method relies critically on a statistical imputation of gene expression profiles, which enables us to break down dependencies between subsets of the parameters. To achieve this, we interpolate experimental points by using a non-parametric method based on Gaussian Processes (GPs).
In this section we briefly review the statistical foundations of GPs;
for a thorough review, the reader is referred to \cite{rasmussenWilliamsBook}. A GP is a (finite or infinite) collection of random variables
any finite subset of which is distributed according to a multivariate normal
distribution. As a random function $f(\mathbf{x})$ can be seen as a collection
of random variables indexed by its input argument, GPs are a natural way of 
describing probability distributions over function spaces. A GP is 
characterised by its \emph{mean function} $\mu(\mathbf{x})$ and 
\emph{covariance function} $k(\mathbf{x},\mathbf{x}')$, a symmetric function
of two variables which has to satisfy the Mercer conditions 
(\cite{rasmussenWilliamsBook}). In formulae, the definition of GP can be written
as\[
f\sim\mathcal{GP}\left(\mu,k\right)\leftrightarrow \left[f\left(\mathbf{x}_1
\right),\ldots,f\left(\mathbf{x}_N\right)\right]\sim\mathcal{N}
\left(\left[\mu\left(\mathbf{x}_1\right),\ldots,\mu\left(\mathbf{x}_N\right)
\right],K\right)\]
for any finite set of inputs $\mathbf{x}_1,\ldots,\mathbf{x}_N$. Here\[
K_{ij}=k\left(\mathbf{x}_i,\mathbf{x}_j\right).\]
The choice of mean and covariance functions is largely determined by the 
problem under consideration. In this paper, we will use a zero mean GP with
MLP (Multi-Linear Perceptron) covariance function \cite{rasmussenWilliamsBook}; this choice is motivated by the ability of the non-stationary MLP covariance to capture saturating behaviours such as those frequently encountered in gene expression data.

Given some observations $\mathbf{y}$ of the function $\mathbf{f}$ at certain input
values $X$, and given a noise model $p(\mathbf{y}\mid\mathbf{f},X)$, one can use 
Bayes' theorem to obtain a posterior over the function values at the inputs
\begin{equation}
p\left(\mathbf{f}\mid \mathbf{y},X,\theta\right)=\frac{p\left(\mathbf{y}\mid 
\mathbf{f},X,\theta\right)p\left(\mathbf{f}\mid X,\theta\right)}{p\left(\mathbf{y}\mid X,\theta\right)}\label{Bayes}\end{equation}
where $\theta$ denotes the parameters of the GP prior, called {\it hyperparameters}. One can 
then obtain a predictive distribution for the function value $f^*$ at a new
input point $\mathbf{x}^*$ by averaging the conditional distribution of 
$p(f^*\mid \mathbf{f})$ under the posterior (\ref{Bayes})
\[p\left({f}^*\mid \mathbf{y},X,\mathbf{x}^*,\theta\right)=\int 
p\left(f^*\mid \mathbf{f},X,\mathbf{x}^*,\theta\right)
p\left(\mathbf{f}\mid \mathbf{y},X,\theta\right)d\mathbf{f}.\]

If the noise model $p(\mathbf{y}\mid \mathbf{f})$ is Gaussian, then one can obtain an analytical expression for the
posterior as all the integrals are Gaussian (\ref{Bayes}). We notice that this analytical property remains even if the variance of the Gaussian noise is different at different input points. A further advantage of the Gaussian noise setting is the ability to obtain a closed form expression for the {\it marginal likelihood} $p(\mathbf{y}\vert\mathbf{x},\theta)$, which then enables straightforward estimation of the hyperparameters (and the observation noise variance) through type II maximum likelihood.

\section{The subsystems approach}
The novelty of our approach lies in splitting the system into modules and then performing parameter estimation for each such subsystem independently. Each subsystem is associated with some of the species of the original system and is responsible for producing parameter estimates and simulated time-series for them. It also has a number of inputs, which are other species that influence its behaviour--- in general, these will be the species that appear in the ODEs for the subsystem's own species.

There are a number of ways in which the decomposition of the complete system can be realised. For the sake of simplicity, we define each subsystem as being associated with a single species, and consider as its inputs the GP interpolation of the time-series of the other species that appear in the ODE for that species's concentration. In this way, a potentially large (autonomous) system involving $N$ species is broken down into $N$ non-autonomous subsystems with a single species; the input to each of these subsystems are the GP interpolations of the chemical species which influence the subsystem in the original large system.

\begin{figure}
\centering{
\begin{tikzpicture}[->, >=latex, auto,node distance=10cm]

\usetikzlibrary{arrows,automata,positioning}

\node[state] (A) {A};
\node[state] (C) [below = 1cm of A] {C};
\node[state] (B) [left = 1cm of C] {B};
\node[state] (D) [right = 1cm of C] {D};
\node[state] (E) [below = 1cm of C] {E};

\path (A) edge node {} (B);
\path (A) edge node {} (C);
\path (A) edge node {} (D);
\path (C) edge node {} (E);
\path (D) edge node {} (E);

\end{tikzpicture}
}
\caption{An example of a system with dependences.}
\label{fig:condInd}
\end{figure}
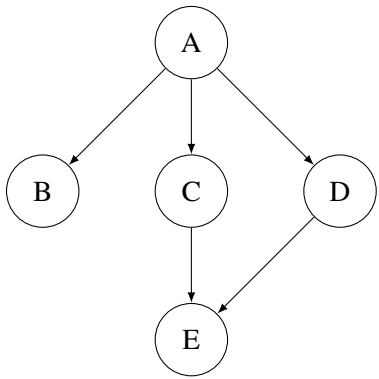

As an illustration, consider the example in Figure~\ref{fig:condInd}, in which the nodes are species and an edge from $x$ to $y$ indicates that the concentration of $y$ depends on that of $x$. We want to reason about the likelihood of a given time-series for species B but, because of the dependence of B on A, we cannot do that unless we know the values of A. However, if we assume that we know the behaviour of A, then B can be treated as an independent part of the system. This is based on the concept of {\it conditional independence}--- knowledge of A makes our belief about B independent of any other species. This is why we use the interpolated time-series of A as input for the B subsystem. Similarly, the E subsystem would require the interpolations of C and D. Essentially, using the interpolation in this way decouples each species from the others, allowing us to follow this modular approach.

There has been a significant amount of research into various approaches to modularisation of biological models (e.g.~\cite{midia}), as well as the conceptual and practical advantages this offers\cite{modularKaltenbachStelling}. In our case, there are two benefits that stand out. Firstly, as each subsystem has a reduced number of parameters directly involved in it, the resulting parameter estimation is performed by searching over a space of lower dimension, and is thus much simpler and more efficient. Secondly, and perhaps more significantly, this procedure is straightforward to parallelise, with minimal synchronisation required. Indeed, each parameter estimation sub-task is independent of the others. Therefore, apart from being simpler than the original, the resulting sub-problems can be solved in parallel.


\section{Experiments}
\subsection{Method}
We estimate the parameters of each subsystem using MCMC, as described previously. We take the prior distribution to be the uniform over an interval $[l,u]$ where $l$ and $u$ are, respectively, the lower and upper limits of each parameter. To calculate the posterior probability of a parameter set $p$, we solve the ODE that corresponds to the subsystem by using these parameter values and the input time-series. We then compare the ODE solution $x = \{x_i(p)\}$ to the input time-series $y=\{y_i\}$ for the subsystem's species to obtain a measure of the likelihood of the data. This is also proportional to the posterior probability of the parameters given the data, $P(p\mid D)$. Specifically, we calculate the square error between the simulated and the input time-series:
$$P(p\mid D) \propto \sum_i(y_i-x_i(p))^2$$
This corresponds to the likelihood of the simulated time-series, assuming a normal distribution with constant variance.

There is an additional component to our method, which involves using a GP-based interpolation method on the original. The benefit of this is two-fold: first, the interpolation smoothes the time-series by removing the noise--- in fact, it estimates the noise level, which does not need to be known {\it a priori}; secondly, it allows us to obtain denser time-series, which we can use as inputs for the different subsystems, as described previously.

In summary, our method is as follows. We begin by performing an interpolation on the original data, obtaining denser time-series. We initially use these interpolated time-series as inputs for the subsystems. For each subsystem, we estimate an optimal set of parameters and calculate the corresponding time-series using these parameters. As mentioned previously, this can be done in parallel. We then gather the calculated time-series and feed them back into the subsystems, replacing the interpolation results. We can repeat this procedure until there is no noticeable difference in the results.

\subsection{Test cases}
We present the performance of our method on two systems. The first is a model of the genetic regulatory network used in the parameter estimation challenge of the DREAM6 contest\cite{DREAM6ParamEstim}. It involves seven species, with ODEs for the concentrations of both proteins and mRNA. It was split into seven subsystems, each encompassing a protein and the corresponding mRNA. The second is a model of a signalling cascade from \cite{Vyshemirsky15032008}, which contains five species and was split into five subsystems as described above.

For the evaluation of our method, we focus on three aspects of the results: we look at how well the interpolation procedure approximates the true data; we examine the fit to the data based on the estimated parameters; and finally, we evaluate the quality of the parameter estimation itself by analysing the results and comparing them to results obtained using state-of-the-art methods.

It is clear that these three aspects are not orthogonal. For example, an inaccurate interpolation will negatively affect the parameter estimation, since the latter uses the former, resulting in an unsatisfactory fit.
 
\subsection{Results and analysis}

\begin{figure}
\subfloat[noiseless measurements] {
\includegraphics[scale=0.3]{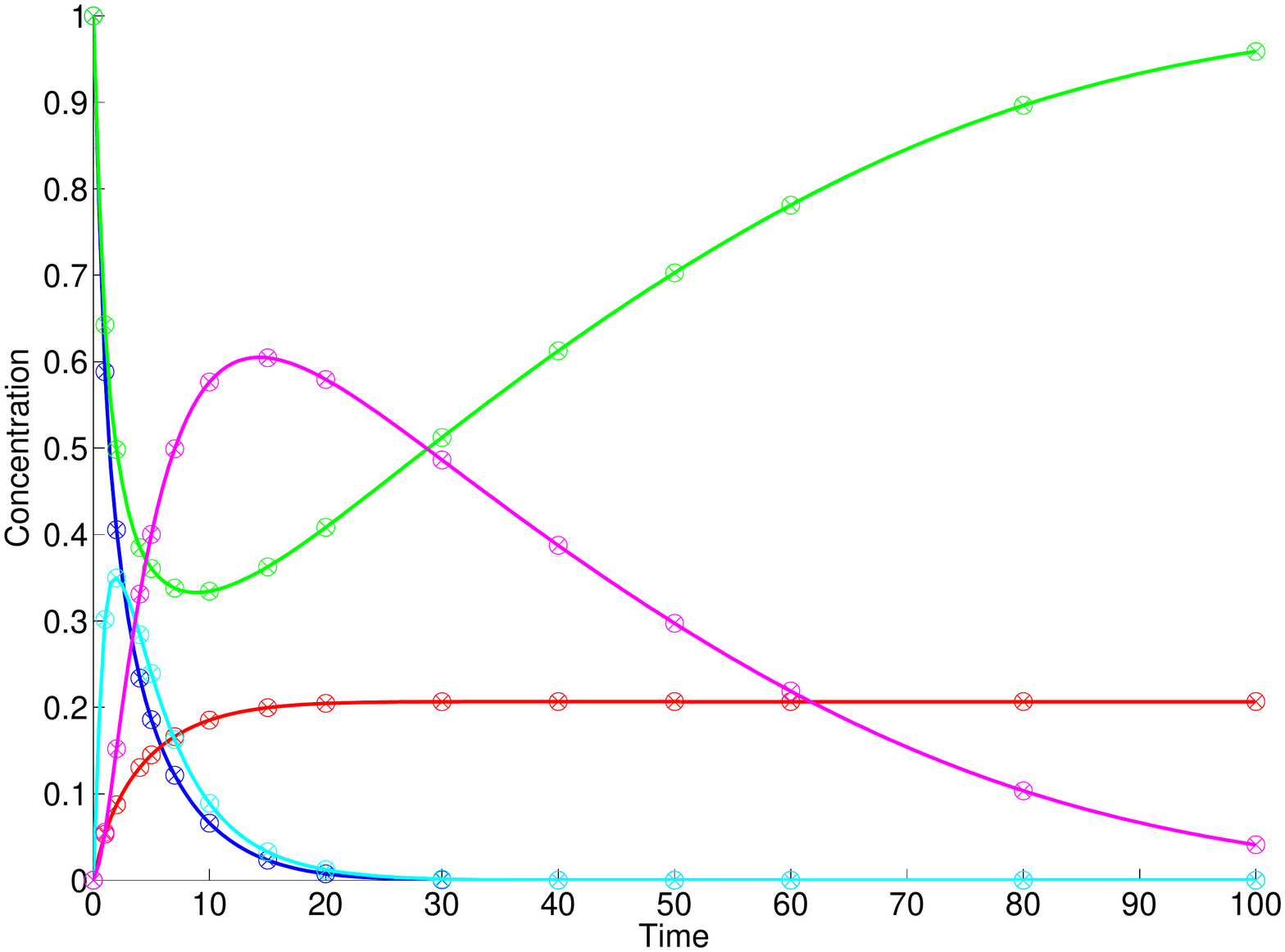}
}
\subfloat[noise with standard deviation 0.5] {
\includegraphics[scale=0.3]{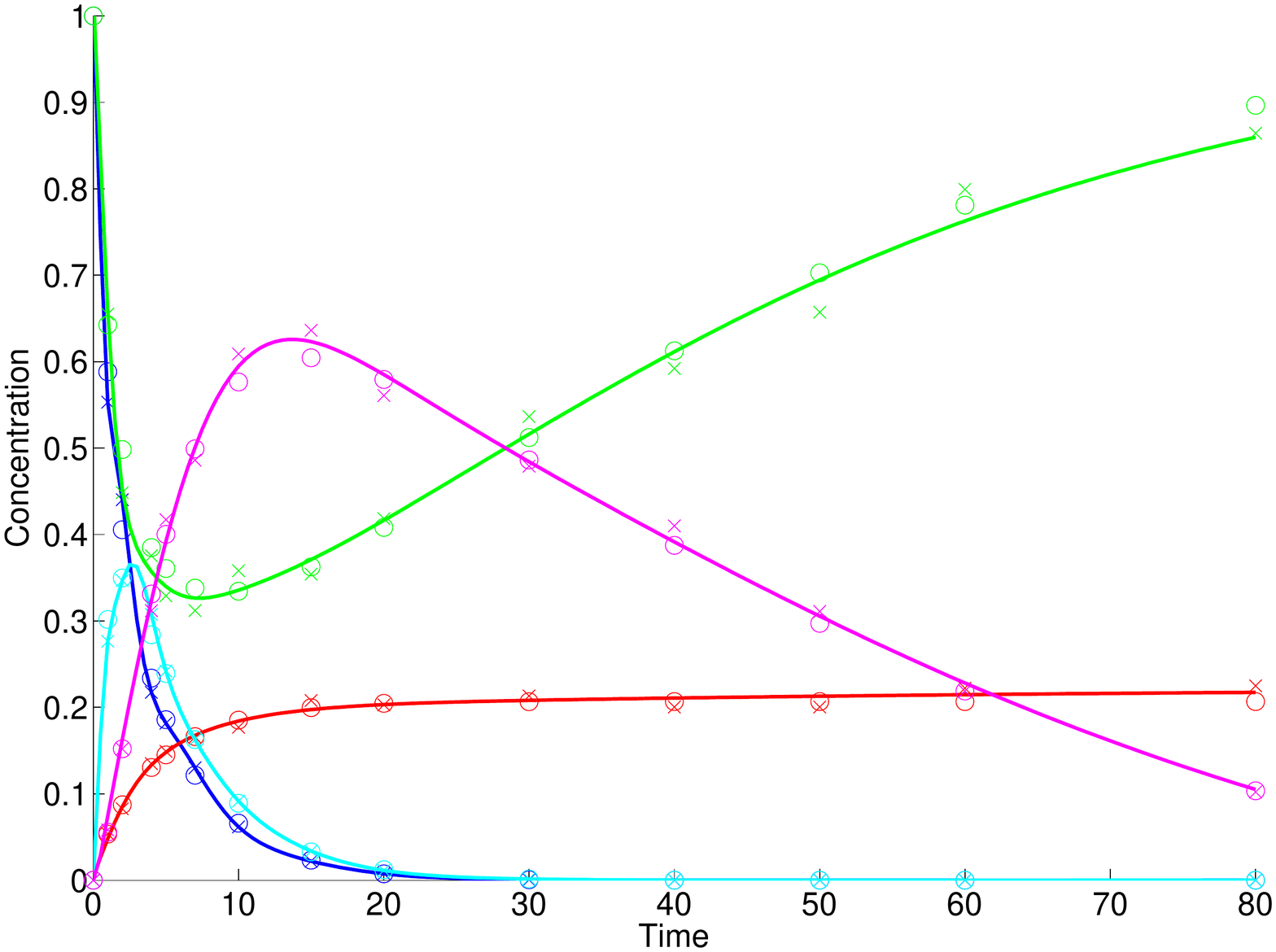}
}
\\
\centering
\subfloat[noise with standard deviation 1] {
\includegraphics[scale=0.3]{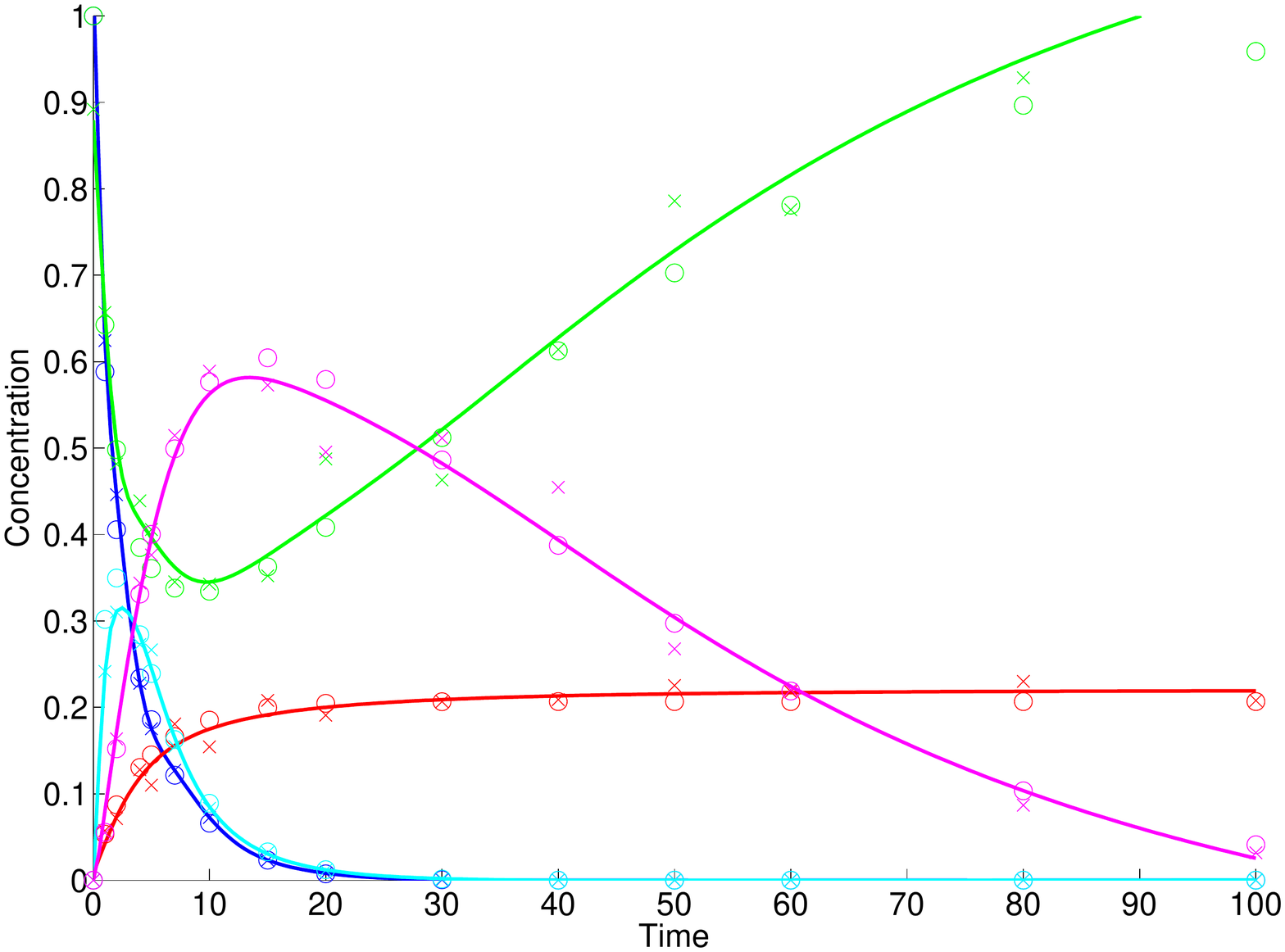}
}
\caption{\label{fig:cascInterp} Signalling cascade model: interpolation (solid line), real data (circles) and noisy measurements (crosses)}
\end{figure}

For the signalling cascade model, Figure \ref{fig:cascInterp} shows the observed data, the interpolated time-series and the real data for different levels of measurement noise. We can see that the GP interpolation is very accurate in approximating the real data--- although, as expected, the accuracy deteriorates as the noise level increases.

\begin{figure}
\subfloat[noiseless measurements] {
\includegraphics[scale=0.3]{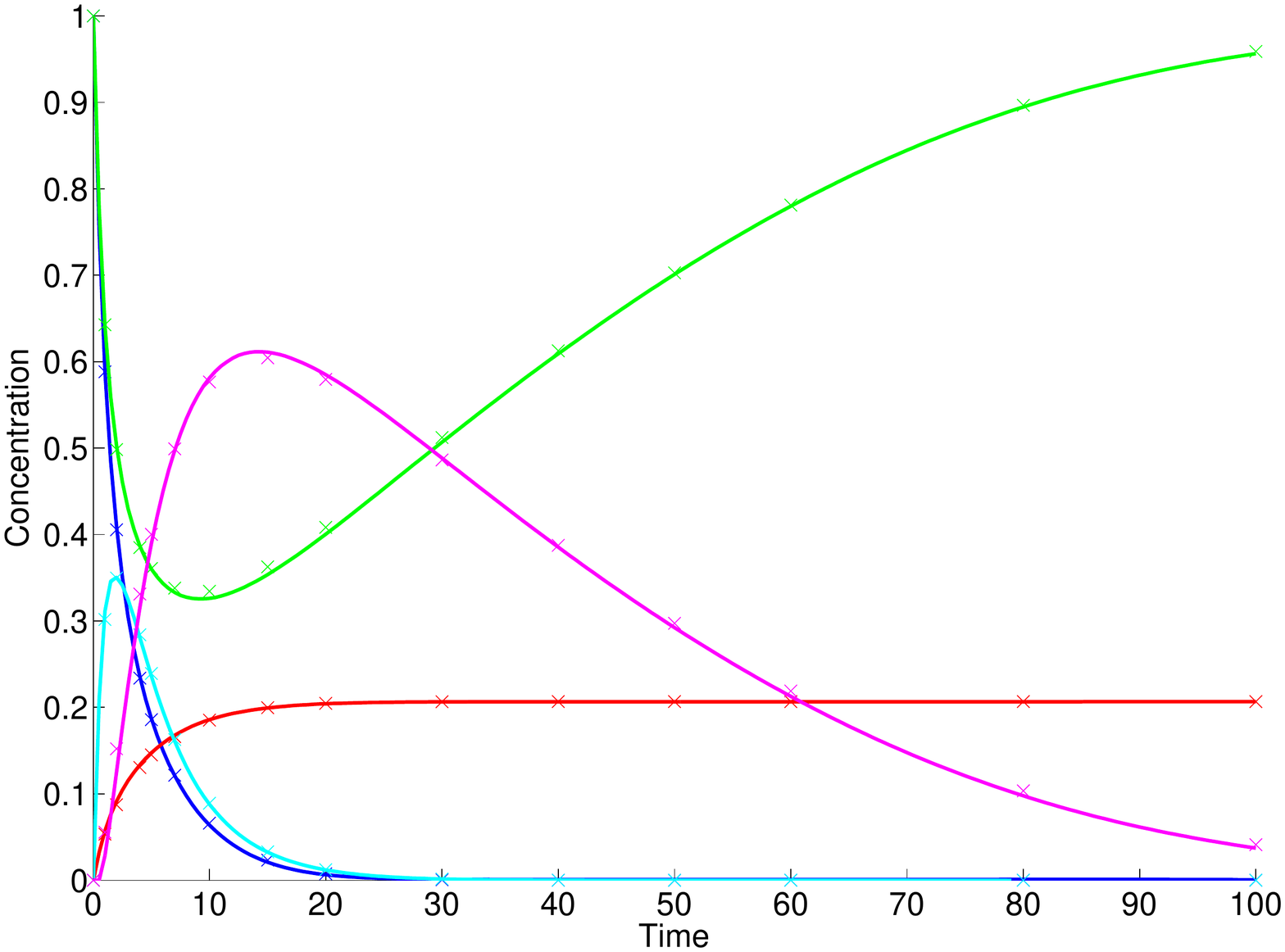}
}
\subfloat[noise with standard deviation 0.5] {
\includegraphics[scale=0.3]{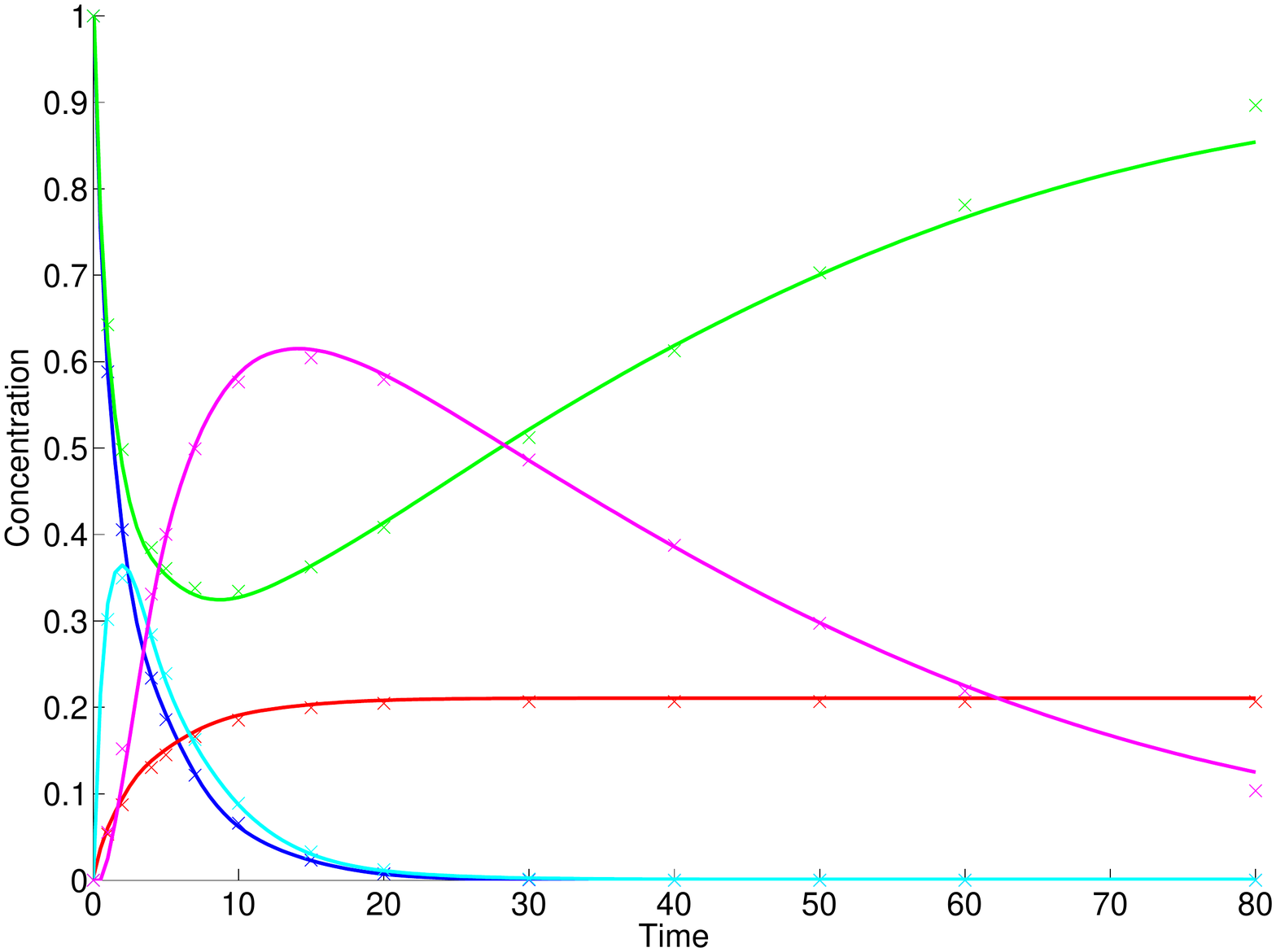}
}
\\
\centering
\subfloat[noise with standard deviation 1] {
\includegraphics[scale=0.3]{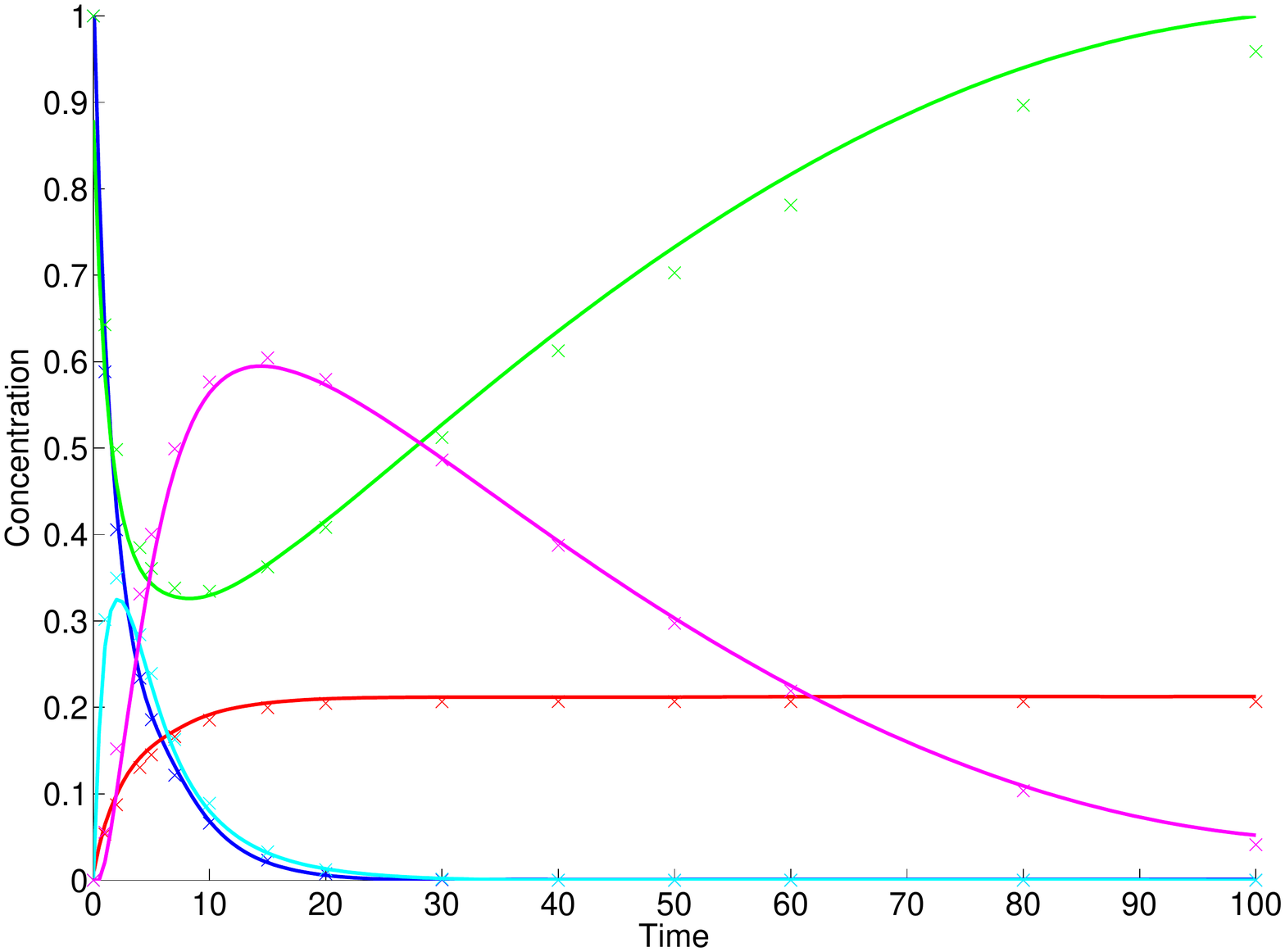}
}
\caption{\label{fig:cascFit} Signalling cascade model: predictions (solid line) and real data (crosses)}
\end{figure}

Figure~\ref{fig:cascFit} presents the predicted output of the model compared to the true data, while Table~\ref{tab:cascEst} shows the results of the parameter estimation. The first thing to note is that we do not obtain unique estimates for each parameter, as they are all involved in more than one ODE (and thus more than one subsystem).We do not have a trivial and fail-safe way to choose between the different estimates, so this remains an open question for our method. However, we should point out that in reality our method does not return a single-point estimate but rather a distribution (the maximum point of which is the value presented in these tables), therefore there is room to ``reconcile" different estimates if we consider them as intervals rather than single values.

\begin{table}
\centering{
\begin{tabular}{l | r | r}
Name & Value & Estimate\\
\hline
$k_1$ & 0.07 & 0.064, 0.172\\
$k_2$ & 0.6 & 0.489, 0.411, 0.345\\
$k_3$ & 0.05 & 0.046, 0.058, 0.025\\
$k_4$ & 0.3 & 0.306, 0.235\\
V & 0.017 & 0.027, 0.036\\
$K_m$ & 0.3 & 0.725, 1.224\\
\end{tabular}
}
\caption{Parameter estimates for the cascade model (multiple values correspond to each parameter's involvement in more than one subsystem)}
\label{tab:cascEst}
\end{table}

\begin{figure}
\subfloat[GP interpolation and fit using estimated parameters] {
\includegraphics[scale=0.2]{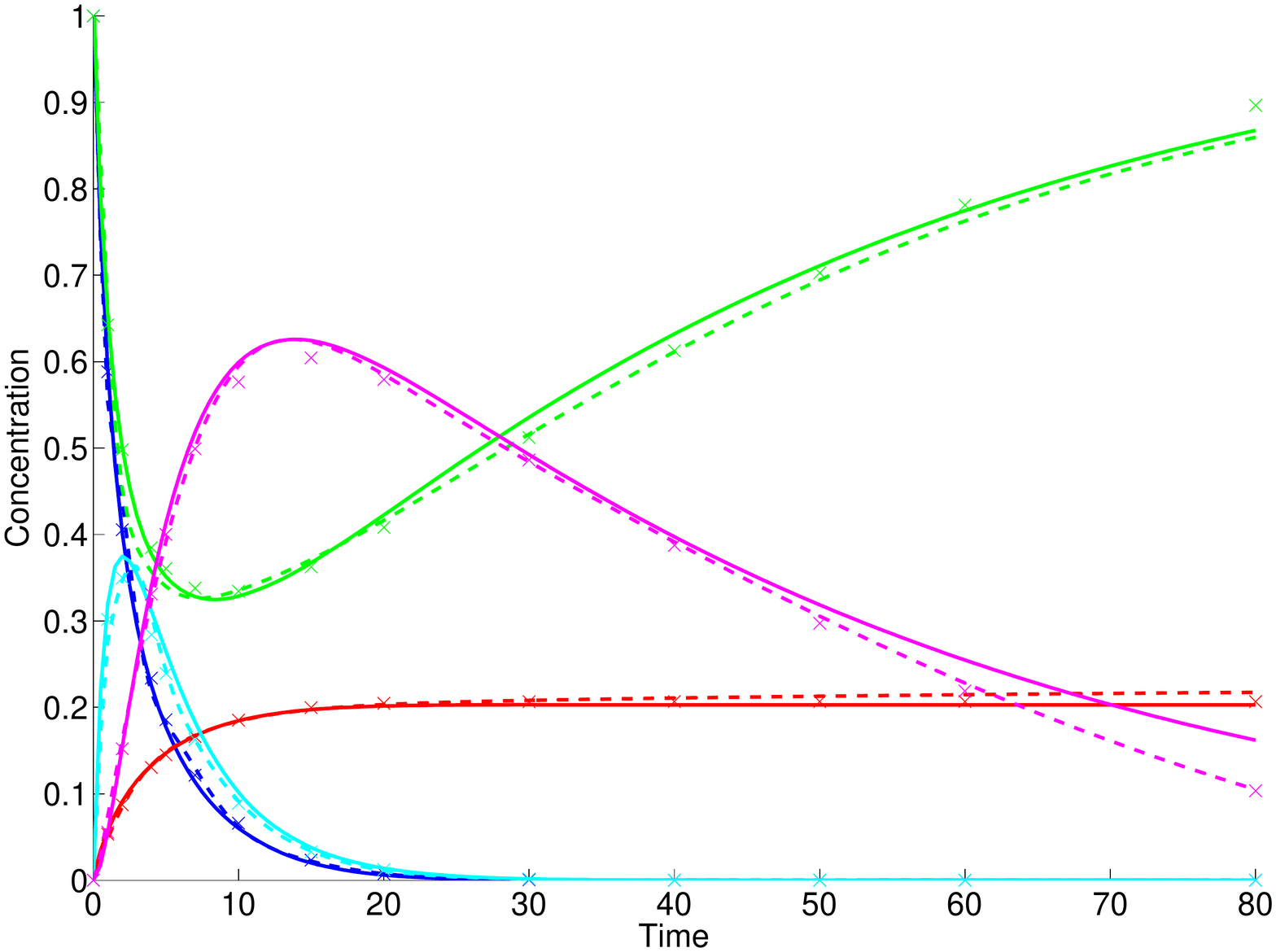}
}
\subfloat[Estimated parameters] {
\raisebox{45pt}{
\begin{tabular}{l | r | r}
Name & Value & Estimate\\
\hline
$k_1$ & 0.07 & 0.348039\\
$k_2$ & 0.6 & 3.0005\\
$k_3$ & 0.05 & 0.160029\\
$k_4$ & 0.3 & 1.41002\\
V & 0.017 & 0.744395\\
$K_m$ & 0.3 & 6.37266
\end{tabular}
}
}
\caption{Results for the cascade model when working with the entire system.}
\label{fig:block}
\end{figure}

Another interesting aspect of the results is seeing how the subsystems approach compares to using MCMC on the system as a whole. As can be seen by comparing Figure \ref{fig:block} to the previous results, the estimates obtained by working on each subsystem separately are much closer to the real values of the parameters, and this is reflected in a better fit to the observed data. This is an encouraging indication that the theoretical benefits of decomposition are also observed in practice.

\begin{figure}
\subfloat[]{
\includegraphics[scale=0.4]{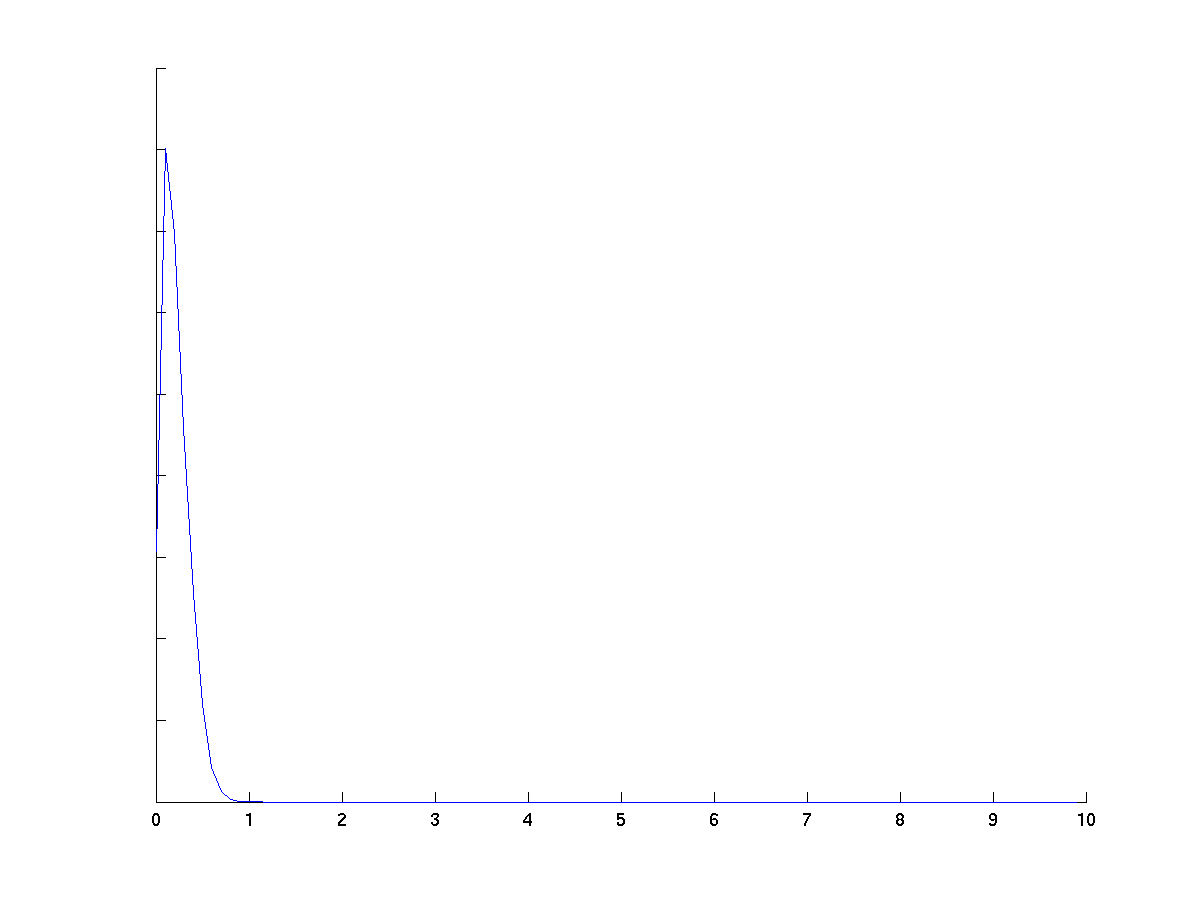}
}
\subfloat[]{
\includegraphics[scale=0.4]{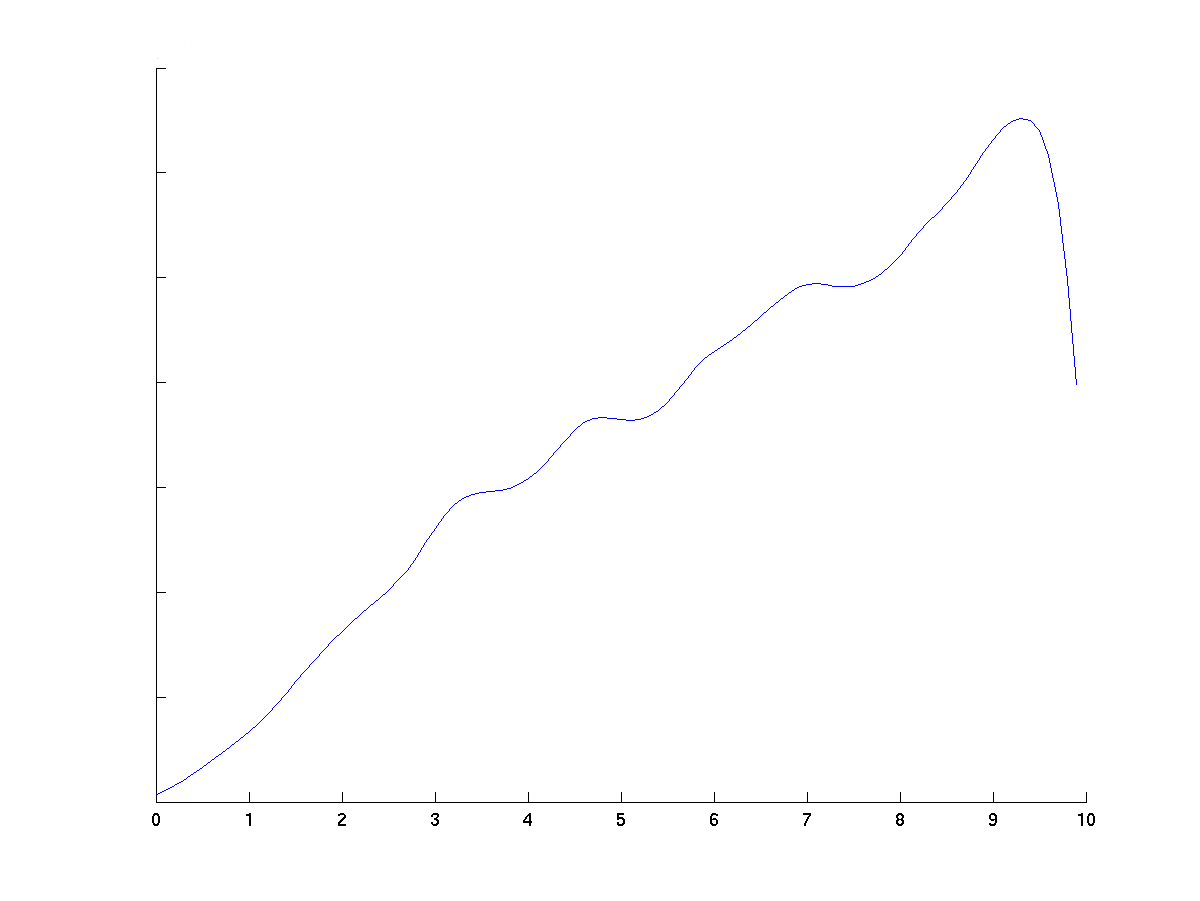}
}
\caption{\label{fig:KDE} Probability density functions for the posterior distributions of the parameters (a) $k_3$ and (b) $V$ in the signalling cascade model, with noise standard deviation 0.5 (similar results were obtained for other noise levels)}
\end{figure}

The results also reveal different confidence levels for the parameter estimates. Figure \ref{fig:KDE} shows an approximation of the posterior distribution for parameters $k_3$ and $V$, as obtained through Kernel Density Estimation on the sampling results. It is clear that the first is quite sharply peaked, whereas in the second the mass is spread over a wider range of values. This shows a higher confidence in the estimate for $k_3$. 

For the gene regulatory network, the performance of our algorithm is more mixed. The time-series for some species are matched very closely to the real data, while for others the fit is not as good. For the sake of brevity, we do not present the plots of all 14 time-series.

This is reflected in the estimated parameter values, and Table~\ref{tab:model2} shows the estimates for each parameter along with the relative error compared to the true values. We can immediately see that the errors span a very wide range. The parameter with the worst estimate is a degradation rate for an mRNA species and it is only involved in one ODE, in a term of the form $pp7\_mrna\_degradation\_rate*pp7\_mrna$. However, the concentration of that mRNA is very low in the measured data, and so the value of the parameter has little to no effect, as the term will always have a value of essentially zero. This makes obtaining an accurate estimate for it impossible using these conditions. We would therefore regard this more as a shortcoming of the data set rather than our method.

\setlength{\tabcolsep}{4pt}

\begin{table}[h]\footnotesize
\subfloat{
\begin{tabular}{l|r|r|r}
 & & Rel. error & Rel. error\\
Parameter & Value & (MCMC) & (COPASI)\\
\hline
pp7\_mrna\_degradation\_rate & 0.217 & 29.23 & 33.28 \\
v3\_h & 0.647 & 13.37 & 2.468 \\
pro2\_strength & 1.614 & 4.278 & 4.376 \\
v1\_Kd & 1.752 & 3.438 & 3.837 \\
pro5\_strength & 0.374 & 2.882 & 6.979 \\
pro7\_strength & 0.984 & 1.342 & 2.463 \\
v3\_Kd & 4.245 & 1.278 & 0.028 \\
pp2\_mrna\_degradation\_rate & 4.928 & 0.955 & 0.506 \\
v4\_h & 7.094 & 0.918 & 0.380 \\
v8\_h & 5.995 & 0.908 & 0.417 \\
v2\_h & 6.838 & 0.907 & 0.364 \\
v1\_h & 9.456 & 0.904 & 0.675 \\
rbs5\_strength & 2.565 & 0.885 & 0.436 \\
v5\_h & 1.871 & 0.875 & 0.777 \\
v2\_Kd & 6.173 & 0.733 & 0.284 \\
v7\_Kd & 1.595 & 0.650 & 5.078 \\
v10\_h & 6.876 & 0.598 & 0.538 \\
v5\_Kd & 9.930 & 0.589 & 0.053 \\
pro4\_strength & 0.653 & 0.581 & 8.001 \\
v10\_Kd & 7.923 & 0.494 & 0.499 \\
p3\_degradation\_rate & 9.948 & 0.490 & 0.413 \\
pp5\_mrna\_degradation\_rate & 3.229 & 0.480 & 0.930 \\
rbs3\_strength & 4.432 & 0.464 & 0.565 \\
v9\_h & 4.523 & 0.453 & 0.060 \\
\end{tabular}
}
\subfloat{
\begin{tabular}{l|r|r|r}
 & & Rel. error & Rel. error\\
Parameter & Value & (MCMC) & (COPASI)\\
\hline
pp6\_mrna\_degradation\_rate & 4.716 & 0.443 & 0.284 \\
pro6\_strength & 6.953 & 0.436 & 0.090 \\
v4\_Kd & 9.674 & 0.424 & 0.397 \\
rbs6\_strength & 1.124 & 0.422 & 0.525 \\
p6\_degradation\_rate & 5.885 & 0.415 & 0.413 \\
rbs4\_strength & 8.968 & 0.366 & 0.317 \\
p4\_degradation\_rate & 4.637 & 0.323 & 0.805 \\
pp4\_mrna\_degradation\_rate & 1.369 & 0.284 & 5.767 \\
rbs2\_strength & 4.266 & 0.272 & 0.782 \\
pp1\_mrna\_degradation\_rate & 3.271 & 0.249 & 0.631 \\
pro1\_strength & 7.530 & 0.222 & 0.079 \\
v9\_Kd & 4.153 & 0.193 & 0.150 \\
p2\_degradation\_rate & 8.921 & 0.183 & 0.105 \\
v8\_Kd & 4.044 & 0.151 & 1.286 \\
rbs1\_strength & 3.449 & 0.149 & 0.788 \\
rbs7\_strength & 9.542 & 0.147 & 0.828 \\
pp3\_mrna\_degradation\_rate & 7.698 & 0.145 & 0.253 \\
p1\_degradation\_rate & 1.403 & 0.137 & 0.280 \\
p7\_degradation\_rate & 5.452 & 0.108 & 0.025 \\
v6\_h & 7.958 & 0.076 & 0.395 \\
pro3\_strength & 4.366 & 0.067 & 0.808 \\
v7\_h & 7.009 & 0.029 & 0.576 \\
p5\_degradation\_rate & 0.672 & 0.026 & 0.017 \\
v6\_Kd & 9.322 & 0.0004 & 0.360 \\
\end{tabular}
}
\caption{Real parameter values and relative errors for the gene regulatory network example, using our method (MCMC) and COPASI.}
\label{tab:model2}
\end{table}

To evaluate these results, we used COPASI\cite{copasi}, a software package for the analysis and simulation of biochemical networks, on the same system and data. The software comes with a number of optimization methods; for the purposes of this comparison, we used {\em simulated annealing}\cite{simAnn}, which explores the search space by attempting to minimize an energy function while also permitting moves towards apparently less optimal solutions, albeit with reduced probability as the search goes on. It is perhaps interesting that some of the other built-in methods could not always produce a result, which we believe was due to numerical computation issues (attempting to invert a singular matrix).

\begin{figure}
\subfloat[]{
\includegraphics[scale=0.25]{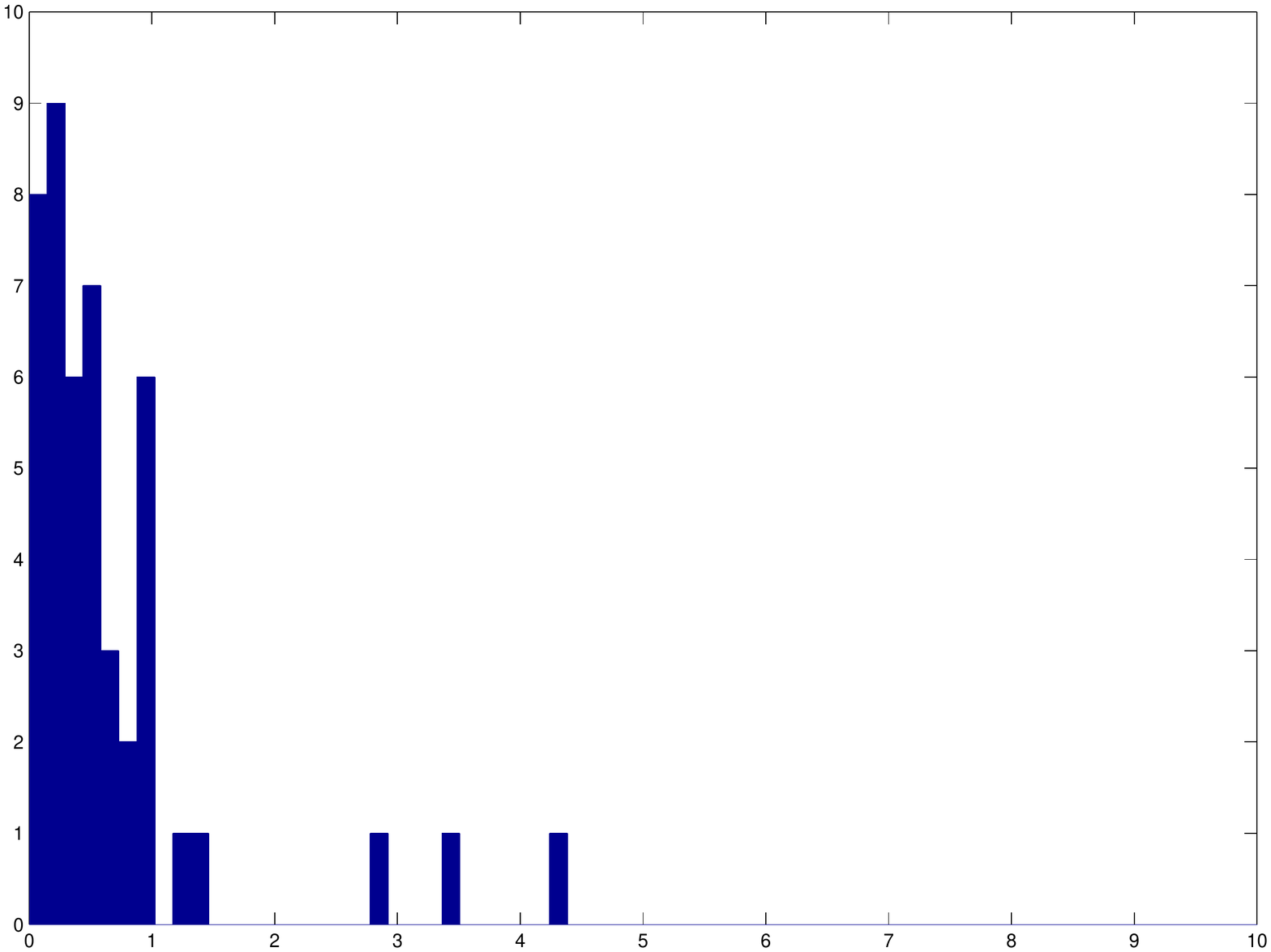}
}
\subfloat[]{
\includegraphics[scale=0.25]{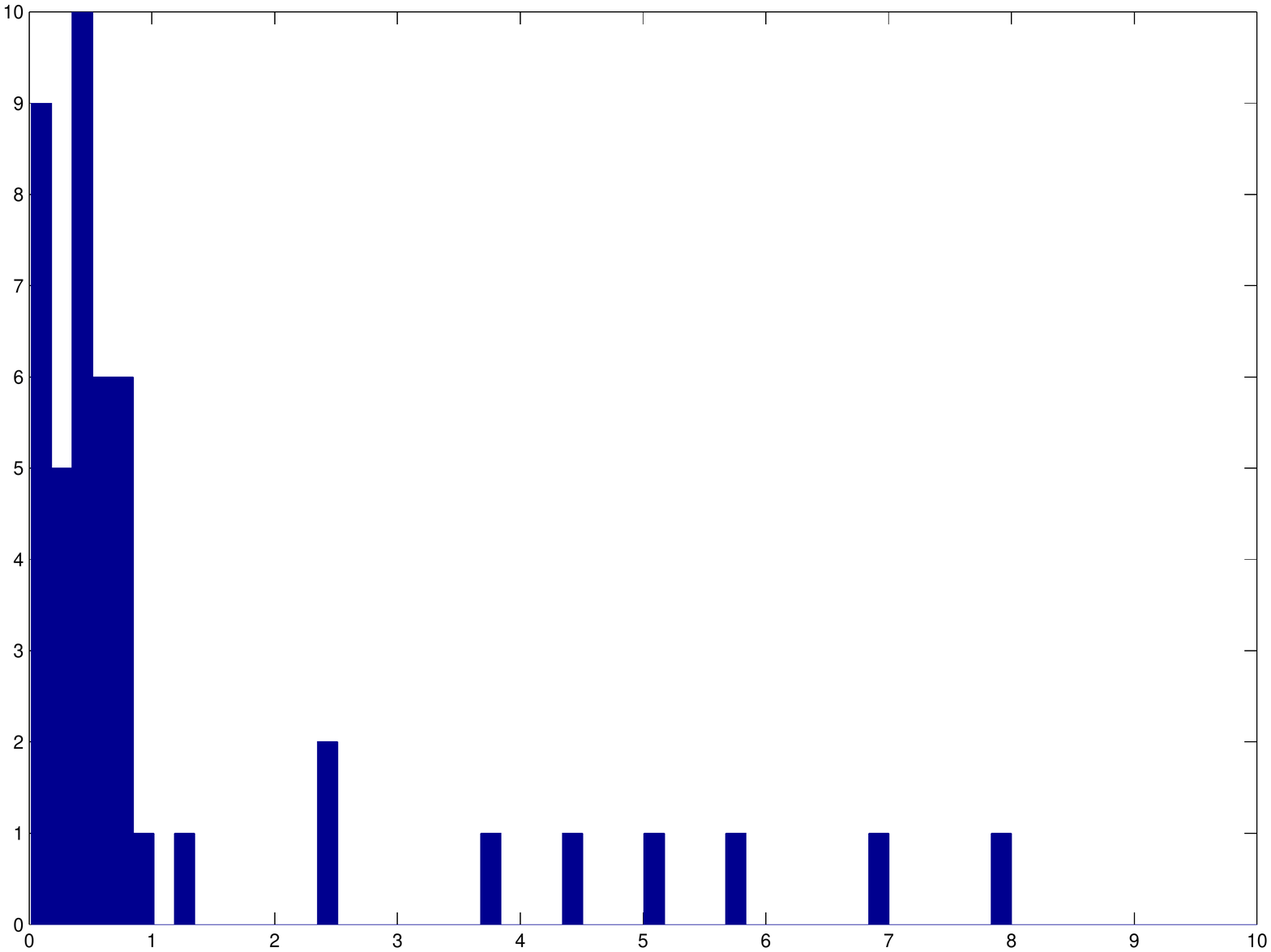}
}
\caption{Distributions of the relative error of the parameter estimates for the gene regulatory network, using (a) our method (2 values excluded) (b) COPASI (1 value excluded).}
\label{fig:hist}
\end{figure}

The parameter estimation results using COPASI are presented in Table~\ref{tab:model2} as well, where we can also observe a wide spread of the errors. Comparing the last two columns, we see that the mean of the errors using our method is slightly better than that using COPASI (1.521 vs 1.862), as is the median (0.448 vs 0.503). For a more general picture, we looked at the distribution of the error values in each case, which is presented in the histograms of figure~\ref{fig:hist}. To present this more clearly, we have excluded the parameters whose estimate had a relative error of more than 10 (i.e. the estimate was an order of magnitude off). This resulted in the exclusion of one parameter for COPASI and of two for our method.

We can see that, using our method, the mass of the errors seems to be more concentrated towards lower values, whereas with COPASI there are more ``outliers" with higher errors. For example, if we look at parameters whose estimate is more than 100\% off the real value, we can find 5 such instances using our method but 9 using COPASI (7 vs 10, if we include the ones excluded from Figure~\ref{fig:hist}), out of a total of 48. This is consistent with the median and mean of the errors being lower with our method, as reported above.

\section{Conclusions and future work}
Parameter estimation remains a central challenge in dynamical modelling of biological systems. While it is most often dealt with in the context of systems of non-linear ODEs, the importance of parameter estimation extends to stochastic and hybrid models, due to the fact that the mean behaviour of a stochastic system is described by a differential equation. In this contribution, we presented a computational approach to speed up parameter estimation in (potentially large) systems of ODEs by using ideas from statistical machine learning. Our preliminary results indicate that the method is competitive with the state of the art, but can achieve significant speed-ups through the ease of parallelisation it entails. Furthermore, the computational resources can be redirected to exploring thoroughly the parameter space of each subsystem, which is often impossible in large systems.

There are several limitations of the current method which clearly point to subsequent developments. Frequently, one is presented with measurements not of a single species in the system, but of a combination of species. For example, one may have access to total protein measurements, but not to measurements of phosphorylated/ unphosphorylated protein levels. A further challenge may arise when the same parameter controls more than one reaction (e.g. in a mass action chemical reaction cascade). A possible solution to this could be to extend the size of the subsystems involved; automatic identification of a minimal size of subsystems however remains problematic.
\section*{Acknowledgements}
The authors would like to thank Dirk Husmeier and Frank Dondelinger for their input and advice.\\
SynthSys (formerly ``Centre for Systems Biology at Edinburgh'') is a Centre for Synthetic and Systems Biology funded by BBSRC and EPSRC, ref. BB/D019621/1.

\bibliographystyle{eptcs}
\bibliography{hsb2012}
\end{document}